\input epsf.sty
\documentstyle[12pt]{article}
\begin{document}
\hsize = 6.0 in
\vsize =11.7 in
\hoffset=0.1 in
\voffset=-0.5 in
\newcommand{\ghat}{{\hat{g}}}
\newcommand{\Rhat}{{\hat{R}}}
\newcommand{\ih}{{i\over\hbar}}
\newcommand{\Scal}{{\cal S}}
\newcommand{\fudge}{{1\over16\pi G}}
\newcommand{\tn}{\mbox{${\tilde n}$}}
\newcommand{\mg}{\mbox{${m_g}^2$}}
\newcommand{\mf}{\mbox{${m_f}^2$}}
\newcommand{\hk}{\mbox{${\hat K}$}}
\newcommand{\vk}{\mbox{${\vec k}^2$}}
\newcommand{\eqletter}{ \hfill (\theequation\alph{letter})}
\newcommand{\gm}{{(\Box+e^2\rho^2)}}
\newcommand{\eql}{\nonumber &\eqletter \cr
                  \addtocounter{letter}{1}}
\newcommand{\be}{\begin{equation}}
\newcommand{\ee}{\end{equation}}
\newcommand{\bea}{\begin{eqnarray}}
\newcommand{\eea}{\end{eqnarray}}
\newcommand{\beal}{\setcounter{letter}{1} \begin{eqnarray}}
\newcommand{\eeal}{\addtocounter{equation}{1} \end{eqnarray}}
\newcommand{\none}{\nonumber \\}
\newcommand{\req}[1]{Eq.(\ref{#1})}
\newcommand{\reqs}[1]{Eqs.(\ref{#1})}
\newcommand{\larrow}{\,\,\,\,\hbox to 30pt{\rightarrowfill}
\,\,\,\,}
\newcommand{\slarrow}{\,\,\,\hbox to 20pt{\rightarrowfill}
\,\,\,}
\newcommand{\half}{{1\over2}}
\newcommand{\bfx}{{\vec{x}}}
\newcommand{\bfy}{{\vec{y}}}
\newcommand{\zfp}{Z_{{FP}}}
\newcommand{\zf}{Z_{{F}}}
\newcommand{\zr}{Z_{{R}}}
\newcommand{\zop}{Z_{{OP}}}
\newcommand{\zekt}{Z_{EKT}}
\newcommand{\phstar}{{\varphi^\dagger}}
\centerline{\bf YANG-MILLS FLOW AND UNIFORMIZATION THEOREMS} 

\bigskip
\begin{center}
S.\@ P.\@ Braham\\
{\it Center for Experimental and Constructive Mathematics\\
Simon Fraser University\\
Burnaby, BC, V5A 1S6 Canada\\}

\medskip
J.\@ Gegenberg\\

{\it Department of Mathematics and Statistics\\
University of New Brunswick\\
Fredericton, NB E3B 4A3 Canada}\\

\end{center}
\bigskip\noindent
{\bf Abstract}

\noindent
We consider a parabolic-like systems of differential 
equations involving geometrical quantities to examine uniformization 
theorems for two- and three-dimensional closed orientable manifolds.  We 
find that in the two-dimensional case there is a simple gauge theoretic 
flow for a connection built from a Riemannian structure, and that the 
convergence of the flow to the fixed points is consistent with the 
Poincare Uniformization Theorem.  We construct a similar system for the
three-dimensional
case.  Here the connection is built from a Riemannian geometry, 
an SO(3) connection and two other 1-form fields which take their values in the 
SO(3) algebra.  The flat connections include the eight homogeneous 
geometries relevant to the three-dimensional uniformization theorem 
conjectured by W. Thurston.  The fixed points of the flow include, besides 
the flat connections (and their local deformations), non-flat solutions of 
the Yang-Mills equations.  These latter ``instanton" configurations may be 
relevant to the fact that generic 3-manifolds do not admit one of the 
homogeneous geometries, but may be decomposed into ``simple 3-manifolds" 
which do.

\bigskip
\noindent 
UNB Technical Report 97-01

\bigskip
\begin{center}
March, 1997
\end{center}
\clearpage

\section{Introduction}

The uniformization theorem in two dimensions is a powerful tool in geometry 
and topology, with applications in physics.  In essence, the theorem states 
that the topology of a closed orientable two-dimensional manifold (a Riemann 
surface) determines which geometries it admits.  In particular, if the 
manifold has handle number zero, it admits the spherical geometry and its 
local deformations; handle number one admits the flat geometry and its local 
deformations; and for handle number two or greater, the admissible geometry is 
that of the hyperbolic plane and its local deformations.  It is important that 
one cannot deform any of these three geometries to obtain one of the other 
two.

This theorem was proved finally around the beginning of the twentieth 
century by H. Poincare \cite{poincare}.  It is a heroic proof, using 
the most sophisticated mathematics of the day.  Unfortunately, the 
classical proof depended heavily on results in complex analysis, and the 
generalization to three or higher dimensions was not obvious.  In fact, 
the three-dimensional analogue of this theorem was first consistently 
formulated only in the late 1970's by W. Thurston and is called Thurston's 
Geometrization Conjecture \cite{thurston78}.  To date, although no 
counterexamples have emerged and it has been shown to hold for very 
large classes of manifolds, the conjecture remains unproved.

Recently, R. Hamilton and B. Chow have constructed a new proof of the
two-dimensional uniformization theorem using techniques not obviously
restricted to that number of dimensions \cite{ham,chow}.  They consider a
one parameter family of Riemannian metrics $g_{\mu\nu}$ 
on an $n$-dimensional smooth 
manifold $M_n$, with the ``flow" governed
by the Ricci curvature tensor $R_{\mu\nu}$:
\be
{\partial g_{\mu\nu}\over\partial t}=-2R_{\mu\nu}+{2\over n}r g_{\mu\nu}.
\ee
In the above $r$ is given by
\be
r:=\left(\int_{M_n}d^nx\sqrt{g}\right)^{-1}\int_{M_n}d^nx\sqrt{g}R, 
\ee
where $d^nx\sqrt{g}$, with $g=\det\left(g_{\mu\nu}\right)$, is the volume 
element on $M_n$.  It is easy to show that ({\it i}.)  
$\partial r/\partial t=0$ 
along the flow; and ({\it ii.}) the fixed points of the 
flow are the ``Einstein 
metrics", satisfying $R_{\mu\nu}=(r/n)g_{\mu\nu}$.  In two dimensions, the 
Einstein metrics have constant curvature, and include all homogeneous 
geometries; in three dimensions, the Einstein metrics include the 
constant curvature, 
but not all the homogeneous geometries; and finally, in four and 
higher dimensions, the Einstein metrics do not have nearly so clear a 
geometric significance as they do in lower dimensions.

The Hamilton/Chow proof of the uniformization theorem in two dimensions 
analyzes the ``Ricci scalar flow"
\be
{\partial R\over\partial t}=\Delta R+R(R-r),
\ee
derived from the flow of the metric above.  For the cases where $R$ is 
non-positive, it is fairly straightforward to show that the flow converges 
to the fixed points with non-positive constant curvature \cite{ham}.  
It took a few more years to analyze the case of $R>0$, since this involves 
a repulsive fixed point \cite{chow}.

Isenberg and Jackson \cite{isenberg} examined the Ricci flow in three
dimensions in order to shed light on Thurston's Geometrization
Conjecture \cite{thurston78}.  Unlike the two-dimensional case, it is
not true in three dimensions that all closed orientable manifolds
admit one of the constant curvature geometries.  Rather, one must
first canonically decompose the given manifold $M$ into its prime
pieces \cite{prime}, obtained by cutting along 2-spheres and gluing
3-balls onto the cuts until one or both of pieces is homeomorphic to
the 3-ball; then cutting along incompressible $T^2$ until the pieces
are either Seifert fiber spaces or contain no embedded incompressible
$T^2$.  Denoting the resulting manifolds by $M_i$, so that
$M=M_1\#M_2\#...$, Thurston then conjectures that the universal
covers $\tilde M_i$ of each $M_i$ admits one and only one of the {\it
eight} homogeneous geometries possible in three dimensions.  
See Table~\ref{manifolds}.  for a list of these eight geometries.  

\begin{table}
$$\begin{array}{|l|l|l|}
\hline
\mbox{Manifold} & \mbox{Isometry Group} &
\mbox{Metric}\\ 
\hline
S^3 & SO(4) & \cos^2{y}\,dx^2+dy^2+(dz-\sin{y} \,dx)^2 \\
E^3 & R^3\times SO(3) & dx^2+dy^2+dz^2 \\
H^3 & PSL(2,C) & dx^2+e^{2x}\left(dy^2+dz^2\right) \\
S^2\times E^1 & \left(Isom(S^2)\times Isom(E^1)\right)^+ & dx^2+dy^2+\sin^2{y}
\, dz^2 \\
H^2\times E^1 & \left(Isom(H^2)\times Isom(E^2)\right)^+ & dx^2+dy^2+e^{2x}dz^2 \\
\widetilde{SL(2,R)} & Isom(H^2)\times R & 
\cosh^2{y} \,dx^2+dy^2+(dz+\sinh{y} \,dx)^2 \\
Nil & Isom(E^2)\times R & dx^2+dy^2+(dz-xdy)^2\\
Sol & Sol\times \left(Z_2\right)^2 & dx^2+e^{-2x}dy^2+e^{2x}dz^2 \\
\hline
\end{array}$$
\caption{The Eight Homogeneous Geometries}\label{manifolds}
\end{table}

The problem with the approach of Isenberg and Jackson is that on the one
hand, there does not seem to be sufficient structure in the Ricci
flow to cope with the neccessity of decomposing the given manifold as
above; and on the other the fixed points are the 
constant curvature geometries, a proper subset of the homogeneous 
geometries.  Hence, one must first cut and
paste the manifold, then flow the geometry on each piece; and one
must look in general for asymptotics, rather than convergence to
fixed points

What we propose here is a one-parameter family of {\it connections}
whose flow converges to flat connections.  In two dimensions, these
flat connections are equivalent to the constant curvature
geometries.  This suggests the possibility of another proof of the
two-dimensional uniformization theorem.  What is important here is
that this flow generalizes to three dimensions such that the fixed
points of the flow are the eight homogeneous geometries plus certain
``instanton" configurations, which describe ``necks" between
three-manifolds.  We believe that this is a promising aproach to
proving Thurston's Geometrization Conjecture.

In Section 2., motivated by the gauge-theoretic formulation of certain 
two-dimensional gravity theories, we will recast the structure equations for 
a constant curvature Riemannian {\it metric} into the form of a flatness 
condition on an appropriate {\it connection}.
  This sets the stage for
constructing a one-parameter family of connections -- the Yang-Mills
flow.  In section 3., the properties of the Yang-Mills flow will be
explicated.  We will show that the fixed points of the flow are
Yang-Mills connections, a subclass of which is the set of flat
connections, and that the latter describe Riemannian metrics of
constant curvature.  In section 4., we analyze the behaviour of the
flow.   For the cases of spherical
and flat Euclidean topologies, we may conclude that the flow
converges to the fixed points.  We present arguments to show that
with initial conditions consistent with regular torsion-free
Riemannian geometries, the flow actually converges to the fixed
points associated with flat connections, and hence to constant
curvature Riemannian geometries.  This leads us to believe that one
may prove the two-dimensional uniformization theorem using the
Yang-Mills flow.  In section 5., we outline a numerical integration
of the Yang-Mills flow.  Finally, in section 6., we will propose a
generalization of the Yang-Mills flow to three dimensions.  We will
conclude by sketching the outline of the form that a proof of the
Thurston Geometrization Conjecture, using this three-dimensional
flow, might take.

\section{Gauge Theory Form of 2D Gravity}

One of the most fruitful areas of research in fundamental physics at the 
moment begins with the reformulation of Einstein's theory of gravity -- the 
general theory of relativity -- as a {\it gauge theory} of the (complexified) 
rotation group SO(3).  This approach, 
pioneered by A. Ashtekar and his collaborators \cite{ash}, has resulted in
some 
progress in constructing a quantum theory of gravity.  This is in part due 
to the fact that the constraints in the theory become  polynomial 
when expressed 
in terms of the above connection and its conjugate momentum; and this in 
turn has allowed for the construction of solutions of the constraints.  Even 
more striking results have been attained by reformulating lower dimensional 
gravity theories as gauge theories.  In particular, three-dimensional 
general relativity can be formulated as a Chern-Simons gauge theory \cite{3d}; 
and some simple two-dimensional gravity theories can be expressed as 
topological field theories of the so-called BF type \cite{it,lineal}.

This suggests that it might be useful to examine the flow of
connections constructed from the Riemannian geometry, rather than the
flow of the metrics themselves.  For one thing, as in the above
theories of gravity, when expressed in connection form, we expect
that the partial differential equations that describe the flow will
be polynomial, unlike the Ricci flow, where the terms involving the
curvature tensor are non-polynomial in the metric.  More importantly,
we note that the connection formulation of lower dimensional gravity
theories are topological field theories, and provide a more direct
route to the global issues that must be addressed.

The fixed points of the Ricci flow are the Einstein spaces.  In two dimensions 
these are just the spaces which have constant curvature Riemannian metrics:  
$R=2k$.  
About ten years ago, Jackiw and Teitelboim (separately) considered the above as 
a toy model of gravity in two dimensions \cite{jt}.  A few years later, at 
least three groups constructed a gauge theory formulation \cite{it}.  
In this formulation, the field equations of the theory required that a 
certain connection over spacetime is flat;  this in turn was equivalent to 
the existence of a constant curvature (pseudo-)Riemannian geometry on the 
spacetime.  In the following, we will construct the connection which has this 
property.   

We now consider Riemannian geometry in the first-order Cartan formalism.  
Instead of the metric tensor $g_{\mu\nu}$ we consider a 
frame-field $e^a$, a set of two 1-form 
fields on $M^2$.  The indices $a,b,..=1,2$.   The metric and frame-fields are related by $g_{\mu\nu}=
\delta_{ab}e^a_\mu e^b_\nu$.  Instead of the Christoffel symbols, we have 
the spin-connection $\omega^a{}_b$, also a set of 1-form fields on $M^2$.  
The spin-connection is skew-symmetric in the indices $a,b$, so in two 
dimensions, there is only one algebraically independent component, which 
we denote simply by $\omega$, defined by 
$\omega^a{}_b=-\omega\epsilon^a{}_b$.  

We can now define a 
connection 1-form field $A$:
\be
A:=e^a P_a + \omega J,\label{eq:conn}
\ee
where $\{P_a,J\}$ generate the Lie algebra:
\be \left[P_a,P_b\right]=k\epsilon_{ab} J;\,\,\, \left[J,P_a\right]=\epsilon_{ab}
 \delta^{bc}P_c,\label{eq:alg}
\ee
with $\delta^{ab}$ the Euclidean metric.  
The algebra generated by $\{P_a,J\}$ is so(3) if $k=+1$,
iso(2) (the ``Poincare algebra") if $k=0$, or so(2,1) if 
$k=-1$.   
If the torsion
\be
T^a:=de^a-\epsilon^a{}_b \omega\wedge e^b=0.
\ee
then the connection $A$ determines a Riemannian geometry.

The curvature 2-form corresponding to the connection $A$ is given by:
\bea
F(A):&=&dA+\half [A,A]\\ \none
&=&T^a P_a +\left(d\omega+{k\over 2}\epsilon_{ab}e^a 
\wedge e^b\right)J\\ \none
&=&T^a P_a -\half\left(R-2k\right)v J\label{eq:curvature}.
\eea
In the above, $v=\half\epsilon_{ab}e^a\wedge e^b$ is the volume element on the 
manifold $M$ induced by the Riemannian metric.

If the connection $A$ is flat, then the curvature $F=0$, and hence 
\bea
T^a&=&0;\\
R&=&2k.
\eea
Hence, a flat connection $A$, with algebra given by $k$, determines a 
Riemannian geometry with constant curvature $2k$, and vice-versa.

\section{Yang-Mills Flow}
In this section, we will describe a one-parameter family of connections, of 
the form given in the previous section.

We start with a two-dimensional manifold $M$ and an admissible Riemannian 
metric $\bar g_{\mu\nu}$  
and a (not necessarily compatible) spin-connection 
$\bar\omega$.  We use this structure to define the 
{\it duals}  
of form fields, e.g. 
\be
*\left(b_\mu dx^\mu\right):=\bar g^{1/2}\epsilon_{\mu\nu}\bar 
g^{\nu\sigma}b_\sigma 
dx^\mu,
\ee
where $\bar g$ is the 
determinant of $\bar g_{\mu\nu}$.  The algebra 
\req{eq:alg} is characterized by the constant $k$.  This is determined from 
the topological structure of $M$ by the Euler number $\chi(M)$:
\be
k:={\chi(M)\over\mid \chi(M)\mid},
\ee
if $\chi(M)\neq 0$, and by 
\be
k=0,
\ee
if $\chi(M)=0$.
In fact, $\chi(M)$ can be computed from 
$\bar g_{\mu\nu}$ by
\be
\chi(M)={1\over4\pi}\int_M d^2x \sqrt{\bar g}\bar 
g^{\mu\nu} \bar R_{\mu\nu} 
,\label{eq:euler}
\ee
where $\bar R_{\mu\nu}$ is the Ricci tensor of $\bar g_{\mu\nu}$.  
The Euler number $\chi(M)$ is related to 
the handle number, or genus $h(M)$ of $M$ by $\chi(M)=2-2h(M)$.

We now flow the connection $A(t)$  
given by 
\be
A(t)=e^a(t)P_a+\omega(t)J.
\ee
The initial values are given by 
\be
A(0)=\bar e^a P_a+\bar\omega J.
\ee

The differential equations that determine the flow are
\be
{\partial A\over\partial t}=-*D_A*F(A).\label{2dflow}
\ee
The dual $*$ and the Lie algebra are determined by the initial fields, 
as discussed above.  Hence $*$ and $\partial/\partial t$ commute.

The parabolic-like structure of the flow is displayed most transparently in 
the equations for the flow of the {\it curvature}.  Using \req{2dflow} and 
the duality in two dimensions of 2-forms with 0-forms, we arrive at
\be
{\partial f\over\partial t}=\Delta_A f,\label{fflow}
\ee
where $f:=*F(A)$ is a Lie algebra valued 0-form equivalent to the curvature; 
and $\Delta_A:=-\left(*D_A*D_A+D_A*D_A*\right)$ is the Laplacian with 
respect to the connection $A(t)$.

The fixed points of the flow are the Yang-Mills connections $A_{ym}$, i.e., 
connections which satisfy the Yang-Mills equations:
\be
-*D_{A_{ym}}*F(A_{ym})=0.\label{ym}
\ee
There are two types of Yang-Mills connections.  The first are flat connections, 
i.e. $F(A)=0$.  If the $e^a$ are non-degenerate, these connections are 
equivalent to the constant curvature Riemannian geometries, as we discussed 
in the last section.  The second type of Yang-Mills connections are 
``instantons," with $F(A)\neq 0$.  In this case, the structure group of the 
connection is reduced to a subgroup which commutes with $f:=*F$ \cite{ym}.

In order to prove a two-dimensional uniformization theorem, we must 
establish the following:

\noindent {\bf Conjecture}:  {\it From an initial connection corresponding to a
sufficiently smooth non-degenerate Riemannian geometry on a
2-manifold $M$ with Euler number $\chi(M)$, the Yang-Mills flow converges to
the
flat connection corresponding to a Riemannian geometry with constant curvature 
having the same sign as $\chi(M)$.}

In the following two sections we will provide analytical and 
numerical evidence for this conjecture.

\section{Convergence of the Yang-Mills Flow:  Analytical Evidence}
There has been some discussion of the properties of Yang-Mills (and
related) flows by mathematicians and mathematical physicists
\cite{ymflow}.  It is clear from this literature, in particular from
the thesis of Rade, that the flow exists and is unique for
short times, at least for the case of non-negative Euler number.
Unfortunately, the situation with regard to the question of the
convergence of flow as $t\to\infty$ is not clear at the moment.  In
the following, we will discuss the question of convergence from an
analytic (but fairly heuristic) perspective.  In the next section,
encouraging results from numerical treatment will be presented.

The Yang-Mills flow resembles the heat equation
\be
{\partial\phi\over\partial t}=\Delta \phi,\label{heat}
\ee
where $\Delta$ is the Laplacian operator with respect to some Riemannian 
structure defined on the manifold $M$ upon which the field $\phi$ takes its 
values.  The existence/uniqueness for short times and the convergence as 
$t\to\infty$ to the ``average" initial data is well-known \cite{heatflow}.
Indeed, it is easy to see that for initial data infinitesimally close to a 
fixed point, the Yang-Mills flow is parabolic.  In general the Yang-Mills 
flow is polynomially non-linear.

We will now discuss the question of convergence for each of the cases 
$k=0$ and $k=+1$.  The $k=-1$ case is the least well-understood at the 
moment, and is under investigation by the authors.

For the case $k=0$ the Lie algebra is ISO(2), which is a semi-direct 
product of the Abelian group SO(2) with the two-parameter group of 
translations.  The Yang-Mills flow itself splits into an SO(2) piece which 
depends only on the $A^2$-component of the ISO(2) connection:   
\be
{\partial A^2\over\partial t}=-*d*dA^2. \label{A2flow}
\ee
Although this is not strictly parabolic, the flow of the corresponding 
dual of the curvature component, $f^2:=*F^2=*dA^2$ is a parabolic system:
\be
{\partial f^2\over\partial t}=\bar\Delta f^2, \label{f2flow}
\ee
where $\bar\Delta$ is the Laplacian with respect to the initial Riemannian 
geometry.   Now the average of the initial curvature component,
\be
\bar F^2:=\int_{M^2}dA^2=0,
\ee
where the last equality follows from the fact the in this case $M^2$ has 
Euler number zero.  Since $f^2$ converges to a constant 0-form, it must 
converge to zero everywhere on $M_2$.  

What we have shown is that the manifold which is topologically $T^2$,
i.e which has Euler number 0, admits a closed 1-form.  This
is a sufficient condition that the manifold admits a Riemanian
geometry with a compatible spin-connection with zero curvature.  To
see this, consider the following:  Let
$\omega$ be a closed 1-form.  
But since
for $T^2$ the space of harmonic 1-forms is two-dimensional, there is 
a harmonic 1-form $\beta$ such that $\omega\wedge\beta\neq 0$. We can
find a chart in which there is a function $p(x)$, such that
$\partial_\mu p(x)=\beta_\mu(x)$.  Now define the 1-forms $e^0,e^1$ in this
chart by $e^0(x)=p(x)\omega(x), e^1(x)=-dp(x)$.  In the chart, the 
volume element $e^0(x)\wedge e^1(x)\neq 0$ by construction, and the 
compatibility condtion $de^a-\epsilon^a{}_b\omega\wedge e^b=0$ since 
$d\omega=0$.

It remains to address the question of which initial conditions, if any, 
determine flows which converge to 
the instantons, with $F(A_\infty)\neq 0$.

In the $k=+1$ case, for which the gauge group is SO(3), we have the
results of Rade \cite{ymflow} wherein in is proved that for compact
simple gauge groups, {\it e.g.} SO(3), the Yang-Mills flow converges
with respect to the Sobolev norm $H^1$ to a Yang-Mills connection.
What remains here is the question of the instantons, as well as
whether the flow converges under stronger smoothness
requirements.\footnote{It is easy to show that, analytically, a round sphere 
with arbitrary radius
will exponentially converge to the sphere with ``correct" radius
$\sqrt 2$, {\it i.e.} so that $R=2$.}

\section{Convergence of the Yang-Mills Flow: 
Experimental Differential Geometry}
The system of coupled partial differential equations 
\req{2dflow} is comprised of
polynomially {\it nonlinear} PDEs, and is therefore potentially quite 
complicated. 
However, the right-hand side of \req{2dflow}
contains second derivatives in
the spatial variables and is therefore much like a diffusive system with
strange convective terms. 
If we 
effectively restrict consideration to
the very high frequency
components of \req{2dflow} by only retaining terms that
contain the highest number of spatial derivatives, we find that
\be
\frac{\partial A}{\partial t} \approx \delta d A,\label{highf}
\ee
where $\delta = -*d*$. The right-hand side of
\req{highf} is not the Laplacian on $M$ defined by
the initial connection, which would be given by
$\Delta = \delta d + d \delta$. Thus, the system does not, even in this
approximation, represent an exact diffusive evolution. However, under
the same conditions for which \req{highf} is valid, we have
\be
\frac{\partial f}{\partial t} \approx \Delta f,\label{fhighf}
\ee
which {\it is} a diffusive equation. It is therefore highly probable
that high frequency spatial perturbations in the curvature
and torsion induced on
$M$ by $A(t)$ are rapidly damped as $t \rightarrow \infty$. We
would therefore expect 
a breakdown of our conjectured 
convergence 
behaviour 
of the flow
only 
if the low frequency modes do not have the appropriate
evolution. These modes are dominated by the nonlinear coupling in
\req{2dflow}.

It seems highly probable that such a system
will not lend itself to easy analytic study. Rather, before 
attempting such an analysis, it seems appropriate that we should
verify the conjectured behaviour as best we can. To do this, we
resort to techniques of {\it experimental mathematics},
wherein we
view the conjectured convergence behaviour as a {\it hypothesis}. 
We will seek experimental evidence for or against the hypothesis.

We can then study the following two questions:
\begin{enumerate}
\item Does $A$ evolve smoothly under \req{2dflow} from smooth initial 
data?
\item Does $A$ converge to a fixed point in an appropriate space\footnote{
In general, we expect convergence within an appropriately normed space,
which we would have to describe. However, since the computer experiments 
use only a fixed number of mesh points
to represent the manifold, all norms are effectively equivalent for purposes
of this section.}
of connections over $M$?
\end{enumerate}

\noindent
If both questions have affirmative answers, we would also like to answer
the concomitant question:
\begin{enumerate}\setcounter{enumi}{2}
\item
Does $F(A) \rightarrow 0$ as $t \rightarrow \infty$?
\end{enumerate}
Should the answers to these questions turn out to be yes, for a reasonably
wide set of initial conditions and topologies, we would have hope that 
further 
study of the flow method ought to be useful in understanding uniformization
theorems. We might further hope that our numerical ``experiments'' would
enable us to observe useful properties of
\req{2dflow} that help suggest ways to prove our conjecture.
To the contrary, should
we find a numerical counterexample to our conjecture, we
could rapidly verify that the conjecture would be false. This is the heart 
of the experimental mathematics concept. 
Given that we conjecture \req{2dflow} to be diffusive and parabolic,
we must first choose a numerical method appropriate to such equations.
It is then an important part of the experimental process to verify
the consistency of the chosen method. 
The explicit forward Euler method is the most naive numerical integration
method appropriate for diffusive parabolic PDEs. It is important to
choose carefully. 
Assume for the moment that the ``convection'' terms 
were to dominate over the ``diffusion''
terms, producing dynamics for $A$ that are far more like hyperbolic
systems. Hyperbolic PDEs are generally unstable
when evolved via naive numerical methods, such as the forward Euler method. 
Furthermore, it would seem
improbable that $A$ would converge if \req{2dflow} was primarily
hyperbolic in nature and if $M$ was compact.

\begin{figure}[ht]
\epsfxsize=3.0in
\centerline{\epsfbox{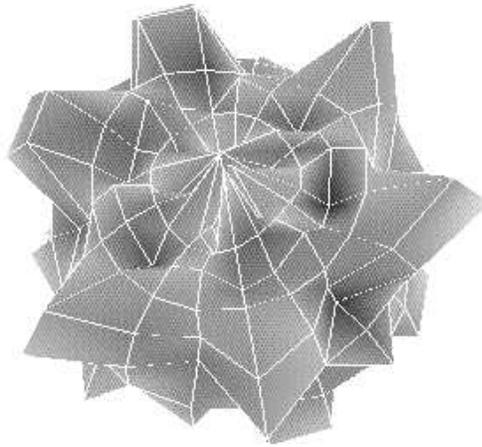}}
\caption{An initial toroidal mesh.}\label{exp71}
\end{figure}

In the Euler method, the iterated value of the connection at time 
$t_{n+1}>t_n$
is taken to be
\be
A_{n+1} = A_n - \kappa (*D_A f(A_n)),\label{FE}
\ee
where $A_n$ is the value of $A$ at a time $t_n:=
t_0 + n \kappa$, $\kappa>0$, and where $A_i$ is 
given on a mesh of discrete points 
approximating 
the manifold $M$. Such
an approximation is shown in Figure~\ref{exp71}, where an initial
connection is represented on a torus. The curvature is represented
by variations of the torus geometry, and the actual torus generated
is only a representation of the state of $A(t)$ and $f(t)$.

We wish
to compare \req{2dflow} to the evolution of a linear system of
the form
\be
\frac{\partial u}{\partial t}=Ku(t),\label{genflow}
\ee
where $u$ is a quantity given on $M$, and $K$ is a linear operator
acting on $u$. The corresponding forward Euler method is
\be
u_{n+1} = u_n + \kappa K u_n,\label{GFE}
\ee
where everything is expressed on a mesh over $M$.
It is well-known that 
\req{GFE} fails to
produce a valid approximation to the solution of the hyperbolic equation
\req{genflow}: the numerically produced solution undergoes
rapid growth in modes that are high in spatial frequency. The approximation
is bad, independently of how small we take $\kappa$ to be. Diffusive
parabolic systems are different in behaviour: for $\kappa$ small enough, the
numerical solution obeys \req{genflow}. 

In view of this comparison, part of our experiment consists of identifying
whether \req{FE} produces a stable approximation (suggesting parabolic
behaviour) or an unstable one (suggesting hyperbolic behavour).

The experimental procedure is as follows: 
the symbolic
system {\it Maple} is used to write part of the computer program,
converting \req{FE} into computer code representing evolution on an
appropriate 
mesh with cell size $h$ and time-step $\kappa$. We then verify the
stability behaviour for $\kappa$ {\it vs}.~$h$. If we observe the 
expected stability behaviour, we can then analyse the resulting evolution
to see if it is consistent with our conjecture.

\begin{figure}[h]
\vspace*{0.5in}
\epsfxsize=2.5in
\centerline{\epsfbox{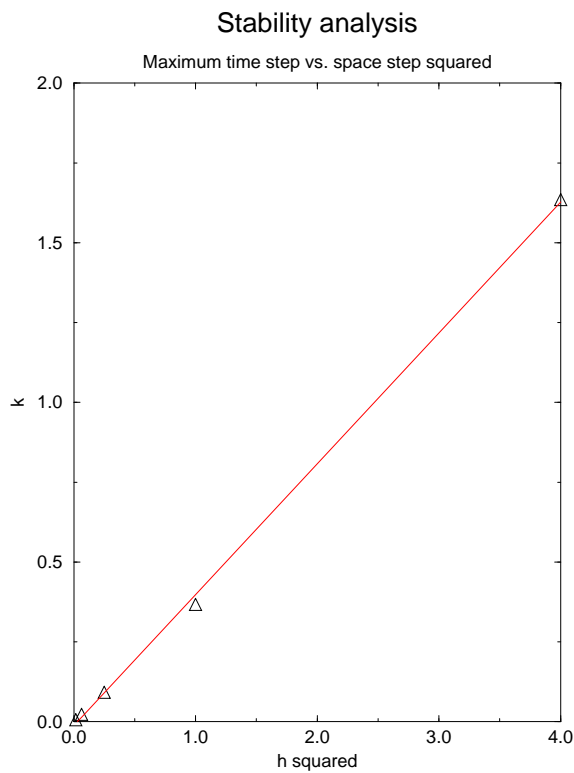}}
\vspace*{0.1in}
\caption{Stability behaviour of the forward Euler method for the
flow equations. Line marks linear fit.}\label{stabfig}
\end{figure}

\begin{figure}[ht]\label{exp715}
\epsfxsize=2.0in
\centerline{\epsfbox{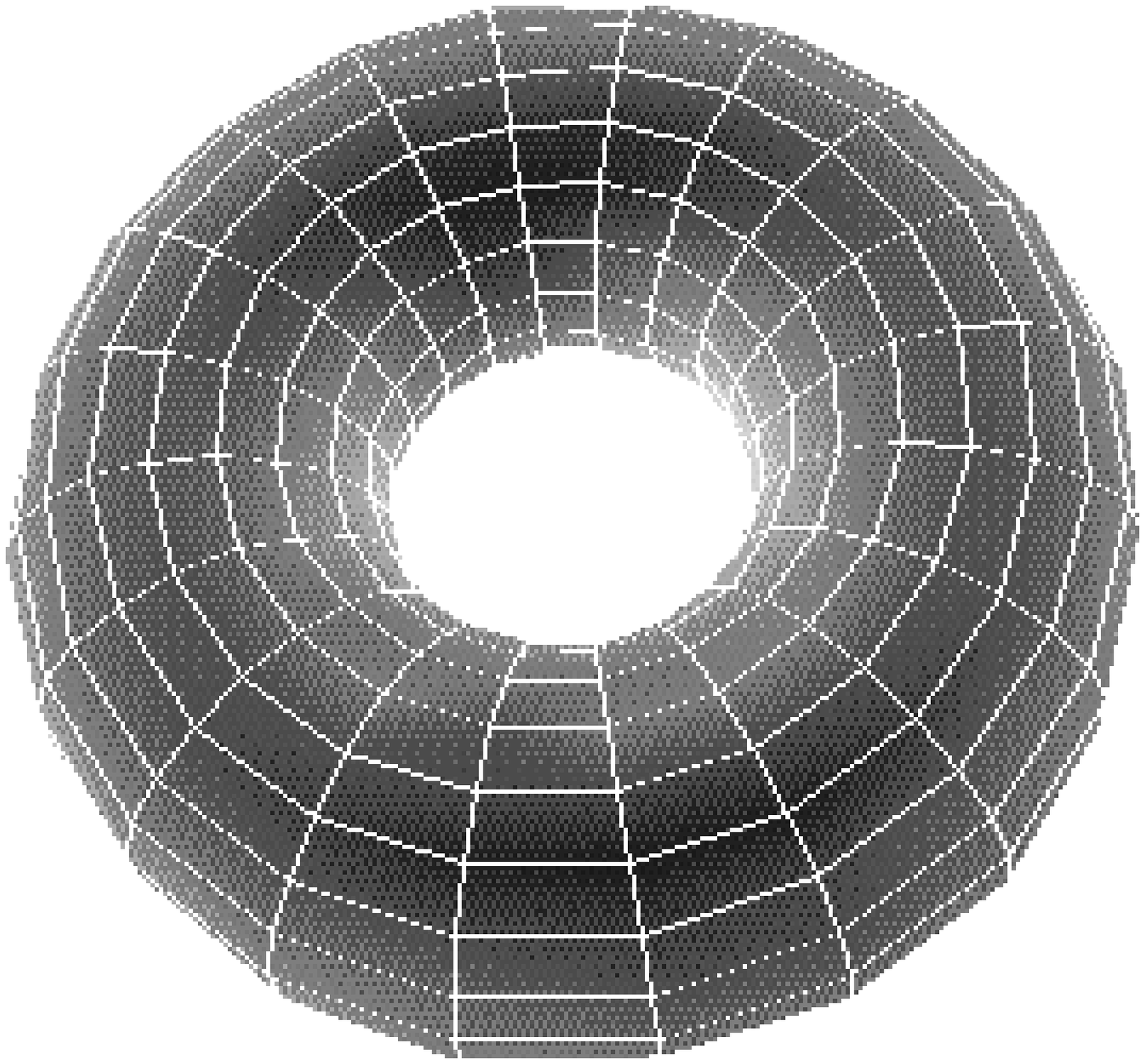}}
\caption{The final configuration, after numerical evolution.}
\end{figure}

Happily, \req{FE} does indeed display, for $k=0$, 
the precise behaviour of a diffusive parabolic PDE. To be precise,
for a toroidal rectangular mesh, 
the numerical stability is observed when $\kappa$ is smaller than a value 
that is of order $h^2$, for a wide range of initial conditions.  One
particular example of this can be seen in Figure~\ref{stabfig}.
This is the anticipated
signature of a diffusive system, whence we have verified one part of
our conjecture, as discussed above.
Furthermore, we obtain evidence that, for the $k=0$ case, all three of our
earlier questions may be answered in the affirmative.
We find that
the numerical evolution of \req{2dflow} under \req{FE} produces
a convergent $A_n$, and that $F(A_n) \rightarrow 0$, as $n \rightarrow
\infty$. This is shown in the final state of the mesh; {\it cf.}
Figure~\ref{exp715}, also drawn on top of the representative torus.
We have therefore found support for our conjectured convergence
behaviour of solutions of \req{2dflow}, at the experimental
level, for $k=0$. 

\begin{figure}[ht]\label{sph3a}
\epsfxsize=2.0in
\centerline{\epsfbox{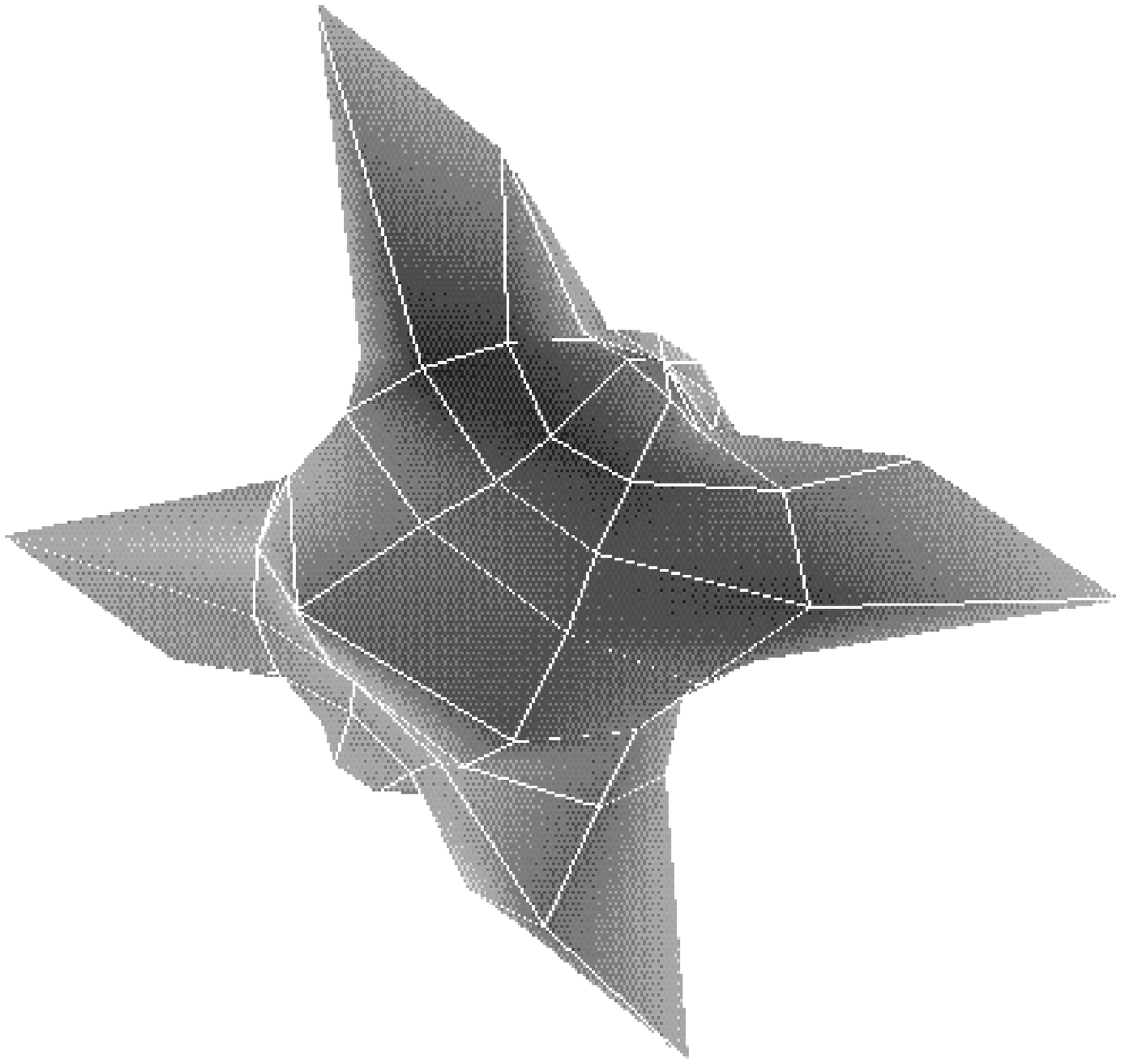}}
\caption{An initial connection for a spherical topology.}
\end{figure}

The conjecture can also be experimentally verified for the
$k=1$ (spherical topology) case. In the spherical case, we construct
an initial connection on a mesh with spherical topology, as shown in
Figure~\ref{sph3a}. The radius from the centre in this image denotes the
curvature radius at that point.
\begin{figure}[hb]\label{sph3b}
\epsfxsize=2.0in
\centerline{\epsfbox{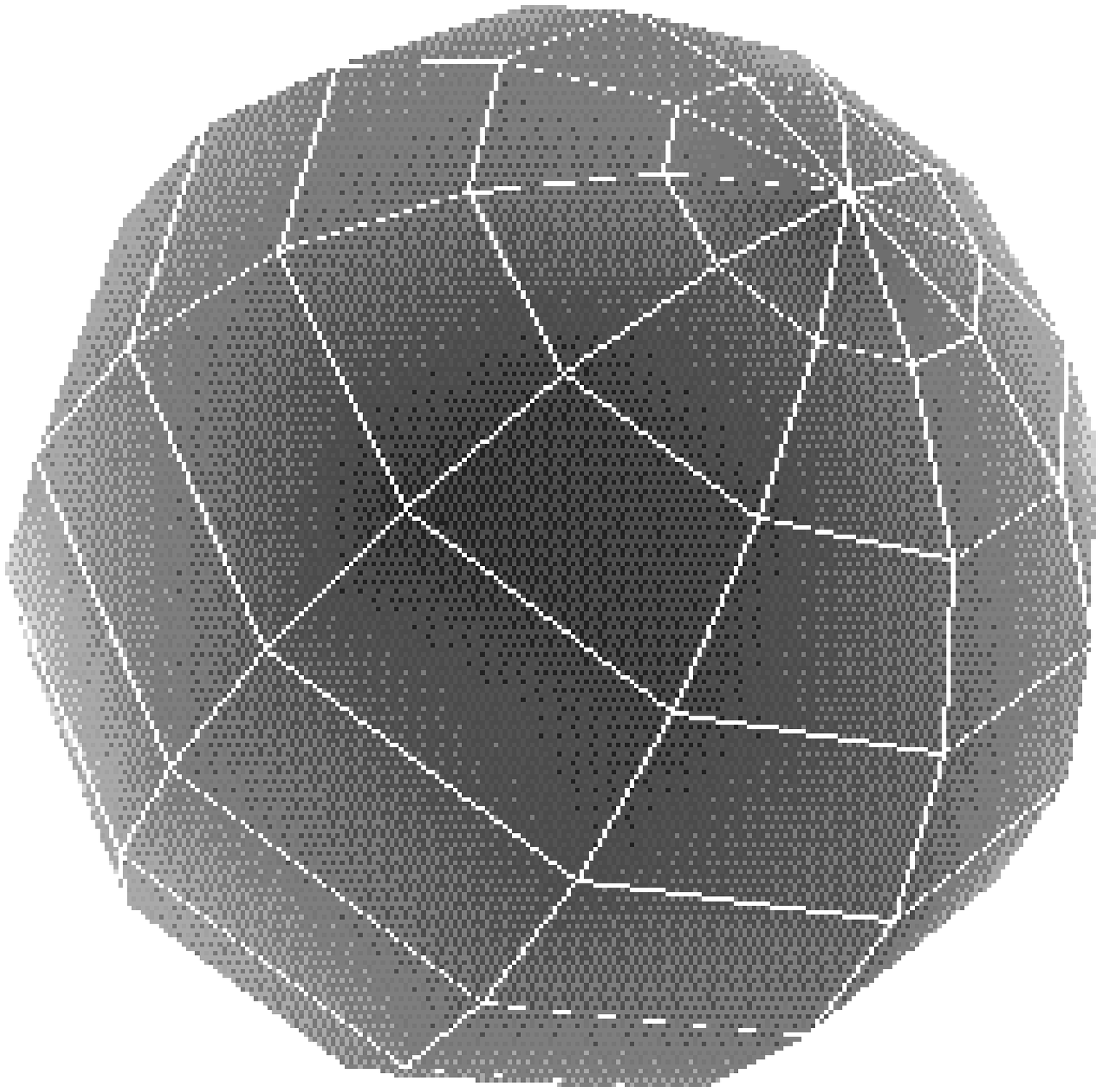}}
\caption{Final, unit sphere, result of evolution.}
\end{figure}
The flow equations can then be evolved. The connection then flows
to one corresponding to a unit 2-sphere, as shown in Figure~\ref{sph3b}.
We thus have experimental confirmation of the conjecture for $k=0$
and $k=1$. 

It remains to verify the conjecture for $k<0$. Unfortunately, 
\req{FE} is not very useful in this case.
The fact that $\kappa$ must be smaller than something of order $h^2$
means that the equations cannot be evolved quickly when we wish to have
$h$ small. For complicated topologies that arise for $k<0$ (particularly
handlebodies, which do not arise in the other cases), we need many cells, 
and $h$ {\it must} be small. Furthermore, it becomes difficult to control
the build up of numerical noise with such a naive, conditionally
stable, method. For $k<0$ ({\it and} the 3-D flow described
in the next section), more sophisticated techniques will be needed.
These techniques will have to deal with the fact that $M$ will
generally need to be covered by more than one coordinate patch, 
more than one gauge patch, will have complicated topology, and
even more nonlinear behaviour. The best practical algorithms seem to be those
using multigrid finite element methods. One of us (SPB)
is currently developing such techniques.
\clearpage

\section{The 3-D Flow}
The gauge-theoretic version of a two-dimensional theory of gravity 
was the starting point for a new approach to proving the two-dimensional 
uniformization theorem.  We emphasize here that we have not completely 
succeeded in constructing the proof; we have only demonstrated its 
plausibility.  Neverthless, this suggests that it might be useful to 
examine gauge-theoretic versions of 3-D gravity theories as possible 
starting points for a proof of the 3-D uniformization theorem conjectured 
by Thurston.

Such theories are well-known.  Einstein gravity (with or without a 
cosmological constant) can be formulated as a Chern-Simons gauge theory 
with gauge group ISO(3) (if the cosmological constant is zero) or 
SO(4) or SO(3,1) if the cosmological constant is positive or negative 
\cite{3d}.  This suggests that we consider a flow of the form
\be
{\partial A\over\partial t}=*D_A*F(A) ,
\ee
where $A$ is the connection 1-form on some 3-manifold for some gauge group 
$G$.  One recovers the Riemannian geometry from the connection 1-form via 
a relation of the form:
\be
A=\omega^aG_a +e^a F_a + ...,
\ee
where the $G_a, F_a$ are generators of $G$, and the $e^a,\omega^a$ can be 
interpreted as a frame field and spin-connection, respectively.  
The $...$ indicates other terms in any additional generators of the 
gauge group $G$.  The idea 
is to choose the gauge group $G$ so that the flat connections (which are a 
subset of the 
fixed points of the flow) 
include at least the {\it eight} homogeneous geometries that occur in 
Thurston's conjecture.  This is not the case if one chooses one of the 
groups (ISO(3), etc.) relevant to Einstein gravity.  
The flat connections for these 
groups determine frame-fields $e^a$ and compatible spin-connections 
$\omega^a$ which have {\it constant curvature}. 

There is a gauge group whose flat connections include the eight 
homogeneous three-dimensional geometries.  The group is the ``doubly 
inhomogenized" group IISO(3).  This group is the semi-direct product of 
the ``Poincare group" ISO(3) with its Lie algebra.  It is a twelve 
parameter non-compact group, whose Lie algebra is
\bea
\left[F^a,G^b\right]&=&\half\epsilon^{abc}F_c;\,\,\,
\left[G^a,G^b\right]=\half\epsilon^{abc}G_c;  \\ 
\none 
\left[G^a,J^b\right]&=&\half\epsilon^{abc}J_c;\,\,\,
\left[G^a,K^b\right]=\half\epsilon^{abc}K_c; \\
\none 
\left[J^a,K^b\right]&=&\half\epsilon^{abc}F_c\,\,\,. \label{pbs}
\eea
The remaining brackets vanish.  The $G^a$ generate the SO(3)
subgroup, while the remaining generators $F^a,J^a,K^a$ behave like
generators of translations, except that the latter two do not
commute.

It was shown in \cite{cargeg} that the Chern-Simons functional with this 
gauge group is equivalent to a three-dimensional theory of gravity 
interacting with topological matter.  This is accomplished by constructing 
the IISO(3) connection as follows:
\be
A=\omega^a G_a +e^a F_a +B^a J_a +C^a K_a.
\ee
If $A$ is flat, i.e. if $F(A)=0$, then by use of the algebra \req{pbs} it 
follows that $\omega^a$ is a flat SO(3) connection, $B^a, C^a$ are 
covariantly constant with respect to $\omega^a$ and $e^a$ satisfies:
\be
D_\omega e^a+\half\epsilon^{abc}B_b\wedge C_c=0, \label{eqmot}
\ee
where $D_\omega$ is the gauge covariant derivative with respect to the 
connection $\omega^a$.

One may now construct a spin-connection which is compatible 
with the frame-field $e^a$, and hence determines a Riemannian geometry.  
In particular, one can show that if the $e^a$ are the frame-fields for 
a homogeneous geometry, then there exist flat connections $\omega^a$ and 
fields $B^a,C^a$ satisfying $D_\omega B^a=0, D_\omega C^a=0$ such that 
$e^a$ satifies \req{eqmot}.  The explicit expressions are shown in
Tables~\ref{es} and~\ref{bcs}, 
where we have taken $\omega^a=0$ for simplicity.

\begin{table}[ht]
$$\begin{array}{|l|l|}
\hline
\mbox{3-manifold}& \left(e^1,e^2,e^3\right) \\
\hline
S^3 &\left(\cos{y}\, dx,dy,dz-\sin{y} \,dx\right) \\
E^3 & \left(dx,dy,dz\right)\\
H^3 & \left(dx,e^xdy,e^xdz\right)\\
S^2\times E^1& (dx,dy,\sin{y} \,dz)\\
H^2\times E^1 & \left(dx,dy,e^xdz\right)\\
\widetilde{SL(2,R)} & (\cosh{y} \,dx,dy,dz+\sinh{y} \,dx)\\
Nil& (dx-zdy,dy,dz)\\
Sol& \left(dx,e^{-x}dy,e^xdz\right)\\
\hline \end{array}$$
\caption{Frame Fields for Homogeneous Geometries}\label{es}
\end{table}
\begin{table}[ht]
$$\begin{array}{|l|l|l|}
\hline
\mbox{3-manifold}&
\left(B^1,B^2,B^3\right)&\left(C_1,C_2, C_3\right)\\
\hline
S^3 &\left(0,dx,0\right)&\left(d(2\sin{y}),
0,
 d(2\cos{y})\right)
\\
E^3 &(0,0,0)&(0,0,0)\\
H^3 &\left(d\left(2e^x\right),0,0\right)& 
(0,d(-z),dy)\\
S^2\times E^1& (d(-2\sin{y}),0,0)&(0,dz,0)\\
H^2\times E^1 & \left(d\left(-2e^x\right),0,0\right)& 
(0,dz,0)\\
\widetilde{SL(2,R)} & (0,dx,0)& 
(d(-2\sinh{y}),0,
 d(2\cosh{y})) \\
Nil& (0,dy,0)& (0,0,d(-2z))\\
Sol &\left(d\left({z-y\over2}\right),d\left(e^x
\right),d\left(-e^{-x}\right)\right)& \left(d\left({z+y\over2}\right),d\left(
2e^x\right),
 d\left(2e^{-x}\right)\right) \\
\hline \end{array}$$
\caption{$B$ and $C$ 1-Form Fields for Homogeneous Geometries}\label{bcs}
\end{table}

It must be remarked here that one may perform an IISO(3) gauge transformation 
on the flat connection equivalent to the homogeneous geometries.  The 
connections are still flat, of course, but in general the $e^a$ are no 
longer frame-fields for a homogeneous geometry.  It is not clear at this 
point how general the gauge-transformed $e^a$ are; though it was shown in 
\cite{cargeg} that for the case of the 3-manifold topology $S^2 \times S^1$, 
all admissible $e^a$ could be obtained from the ``trivial" configuration 
$\omega^a=e^a=B^a=C^a=0$ by a gauge transformation.

If it could be shown that an arbitrary IISO(3) connection on the 3-manifold 
flowed to one of the flat connections, then it would follow that the 
3-manifold would admit the {\it homogeneous} representative of the gauge 
orbit.  This is what we expect of a proof of the uniformization conjecture 
for 3-manifolds.

In the two-dimensional Yang-Mills flow, the fixed points of the system are the 
Yang-Mills connections.  However, it seems to be the case that regular 
initial conditions will flow to the appropriate subclass of flat connections. 
The reason for this is that in two dimensions, the non-flat Yang-Mills 
connections are reducible.  This is not the case in three dimensions, since 
the duals of the curvature 2-forms are 1-forms, and hence {\it not} 
gauge parameters.  Hence we conjecture that in three dimensions some 
regular connections, built from non-degenerate Riemannian metrics, will flow 
to instanton fixed points, not flat fixed points.  This might be related to 
a fundamental difference between the two- and three-dimensional cases.  
Indeed, as we 
discussed in the Introduction, in contrast to the two-dimensional 
case, a given closed orientable 3-manifold does not in general admit a 
homogeneous Riemannian geometry.  Thurston's conjecture is that any 3-manifold 
admits a canonical decomposition into the connected sum of 
{\it simple manifolds}, i.e., prime manifolds 
which either have no incompressible embedded $T^2$ or are Seifert fiber
spaces; 
and each simple manifold in turn admits one and only one of the eight 
homogeneous geometries.  If the flow converged to fixed points which were 
flat connections, then in general it would have to be singular when the 
manifold was not simple.

However, consider the flow of the curvature as above.  
The fixed points are the flat connections
$F(A)=0$ {\it and} the Yang-Mills instantons which satisfy
$*D_A*F(A)= 0, \,\, F(A)\neq 0$.  It would be interesting to examine
IISO(3) connections on 3-manifolds to see if the Yang-Mills equations have 
non-flat solutions when the manifold is simple.  If it turns out that simple 
3-manifolds admit flat solutions of the Yang-Mills equations, then the 
next question to examine would be the structure of 3-manifolds which 
consist of two simple manifolds joined by a ``neck".  If Thurston's 
conjecture is correct, then the following scenario should hold:  
the manifold-with-neck would presumably admit 
an instanton 
which was asymptotically flat at the ends of the neck.  In general, 
non-simple 3-manifolds would {\it not} admit 
non-trivial flat IISO(3) connections.  Work in this 
direction is underway by the authors.

\section*{Acknowledgments}
The authors thank Eric Woolgar for many useful discussions and for
his initial collaboration in this work.  They also thank Steve Boyer
and Jim Isenberg for helpful discussions.

\end{document}